\documentclass[aps, prd, showpacs, twocolumn, superscriptaddress, groupedaddress, floatfix]{revtex4-2}

\usepackage{graphicx}
\usepackage{yfonts}
\usepackage{float}
\usepackage{amsmath}
\usepackage{amssymb}
\usepackage{amsfonts}
\usepackage{amsthm}
\usepackage{mathtools}
\usepackage{derivative}
\usepackage{dcolumn}
\usepackage{listings}
\usepackage{subfigure, rotating, bm, array}
\usepackage[utf8]{inputenc}
\usepackage[english]{babel}
\usepackage[pagebackref=false, colorlinks=true]{hyperref}
\hypersetup{
linkcolor=blue,     
citecolor=blue,     
urlcolor=blue}      

\usepackage{float}

\begin{document}
\title{Gravitational collapse of matter in the presence of non-minimally coupled Quintessence and Phantom-like scalar fields}
\author{Priyanka Saha}
\email{priyankas21@iitk.ac.in}
\affiliation{Department of Physics, Indian Institute of Technology, Kanpur,
Kanpur-208016, India}
\author{Dipanjan Dey}
\email{deydipanjan7@gmail.com}
\affiliation{Department of Mathematics and Statistics, Dalhousie University, Halifax, Nova Scotia, Canada B3H 3J5.}
\author{Kaushik Bhattacharya}
\email{kaushikb@iitk.ac.in}
\affiliation{Department of Physics, Indian Institute of Technology, Kanpur,
Kanpur-208016,India}
\begin{abstract}
This paper explores the evolution of the over-dense region of dark matter in the presence of a non-minimally coupled scalar field which is used to model quintessence and phantom-like dark energy. We focus on algebraic coupling, where the interaction Lagrangian is independent of the derivatives of the scalar field. To make our model more relativistic, like the minimal coupling scenario we studied earlier, we consider a spacetime structure that is internally closed Friedmann-Lemaitre-Robertson-
Walker (FLRW) spacetime and externally the generalized Vaidya spacetime. This structure allows non-zero matter flux at the boundary of the over-dense region. Our investigation reveals that an increment of the coupling strength causes dark energy to cluster with dark matter at a certain cosmological scale where the influence of dark energy cannot be ignored. This phenomenon arises from the specific nature of the non-minimal coupling considered in this paper. While the evolution of matter's energy density remains unchanged, the scalar field's Klein-Gordon equation is modified, causing dark energy to deviate from its homogeneous state and cluster with dark matter. Similar to minimal coupling scenarios, closed spherical regions do not collapse within certain parameter ranges, exhibiting eternal expansion within the spatially flat FLRW spacetime—acting as voids with decreasing matter density. The study extends our understanding of the cosmological scenarios where the virialization of the over-dense regions of dark matter is influenced by the non-minimally coupled dark energy.
\end{abstract}
\maketitle

\section{Introduction}
The exploration of the emergence of structures in a homogeneous and isotropic universe is an intriguing topic in astrophysics and cosmology, rooted in linear perturbation theory in cosmology \cite{Jeans:1902fpv,Moradpour:2019wpj}. As we know, after recombination, perturbation modes deviate from linearity, serving as the pillars for future structural formations. Before entering into the non-linear regime, phenomena such as Jeans instability play an important role in growing the primordial perturbations \cite{Jeans:1902fpv,Moradpour:2019wpj}.
Gravitationally bound structures, from galaxy clusters to smaller scales, are believed to originate from nonlinear instabilities, with the dark matter sector playing a central role \cite{Paddy, Primack, Dey1, Dey3}. Since dark matter seems to be weekly interacting, it decouples from the `primordial matter soup' long before baryonic matter, initiating a collapse to form structures. Consequently, these over-dense regions are conventionally believed to predominantly consist of dark matter. When the baryonic matter decouples, it accumulates inside the over-dense regions of dark matter and eventually forms galaxies, galaxy clusters, etc. 

As an initial approximation, the primordial over-dense patches of dark matter are considered to be spherically symmetric, and their evolution is commonly modeled using the `top-hat collapse model'  \cite{Gunn}. In this model, these over-dense regions are described using a spherically symmetric closed Friedmann-Lemaitre-Robertson-Walker (FLRW) metric. On the cosmological scale, the universe is nearly flat, and therefore, the background of the over-dense regions is described by a spatially flat FLRW metric.
According to the top-hat collapse model, the over-dense regions of dark matter initially expand in an isotropic, homogeneous manner along with the background flat FLRW spacetime. The fluid within these spherically symmetric over-dense regions is considered to be homogeneous and pressure-less. Eventually, the dynamics of these over-dense regions detach from the background cosmic expansion, and these regions begin to behave like sub-universes.  These over-dense sub-universes start to collapse under gravity after reaching a turnaround.
However, within general relativity (GR), it is well-known that the outcome of a collapse involving a homogeneous pressure-less fluid (dust) is always a black hole  \cite{Oppenheimer:1939ue}. Consequently, using the top-hat collapse model alone, it is impossible to relativistically obtain desired small-scale equilibrium configurations as the eventual end-states of the gravitational collapse of over-dense regions. To address this limitation, Newtonian virialization techniques are employed to obtain such equilibrium states, providing an explanation for galaxy formation.

Virialization is a process wherein a system of $N$ particles attains an equilibrium state at a large system time, given by
\begin{equation} \label{eq:vir1}
\langle T\rangle_{\tau\rightarrow\infty} = -\frac12\Bigg\langle \sum_{i=1}^{N}{\bf F}_i\cdot{\bf r}_i  \Bigg\rangle_{\tau\rightarrow\infty},
\end{equation}
where ${\bf F}_i$ is the net force on the $i$th particle, ${\bf r}_i$ is its position, $T$ is the total kinetic energy of the system, and the system reaches the virialized state over a large system time $\tau$. The angular bracket denotes time averaging. If the particles in such a system interact solely gravitationally, the system achieves its final equilibrium state when the following condition is met:
\begin{equation}
\langle T\rangle=-\frac{1}{2} \langle V_T \rangle,
\end{equation}
where $V_T$ is the total gravitational potential of the system. Four processes primarily govern the virialization of a collapsing system, consisting only of gravitationally interacting particles: violent relaxation, phase mixing, chaotic mixing, and Landau damping \cite{Lynden-Bell67, Merritt99}. Using the principle of energy conservation, one can establish that spherically symmetric over-densities undergo virialization when $\eta= \frac{R_{vir}}{R_{max}}=0.5$, where $R_{vir}$ and $R_{max}$ are the physical radius of the over-dense region at the virialization time $t_{vir}$ and the turnaround time $t_{max}$, respectively. One can show that the time $t_{vir}$ is equal to $1.81$ times the time $t_{max}$ if the over-dense region is modeled by closed FLRW spacetime.

In the top-hat collapse model, the aforementioned virialization argument is invoked to stabilize a collapsing system. As discussed previously, in this model, dark matter is considered homogeneous and dust-like throughout the evolution of the over-dense regions, mainly because such a fluid can satisfactorily explain the large-scale structure of our universe. Models in which dark matter is regarded as pressureless and non-relativistic are known as cold dark matter (CDM) models \cite{BullockBoylan-Kolchin17, Weinberg+15}.
Conventionally, the role of the cosmological constant, $\Lambda$, in the structure formation process is often overlooked. However, some authors have attempted to integrate the effects of $\Lambda$ into the gravitational collapse process \cite{Garfinkle:1990jr,Deshingkar:2000hd,Sharif:2006bp,Markovic:1999di}. Traditional $\Lambda$CDM models face inherent challenges \cite{DelPopolo:2016emo,Perivolaropoulos:2021jda}. As a response to these challenges, dynamical dark energy models based on scalar fields have been introduced. One frequently used scalar field in this paradigm is the quintessence field. Additionally, phantom-like scalar fields, characterized by a negative kinetic term, are also employed to model dark energy \cite{Bamba:2012cp,Caldwell:1999ew,Carroll:2003st,Copeland:2006wr,Caldwell:2003vq}.

At a certain cosmological scale, dark energy may have a non-zero contribution to the total gravitational potential $V_T$ of the over-dense region of dark matter. $V_T$ of the over-dense region in a two-fluid system consisting of dark matter (DM) and dark energy (DE) is given by \cite{Maor:2005hq}:
\begin{eqnarray}
V_T &=& \frac{1}{2} \int_v \rho_{DM}\phi_{DM} \,dv + \frac{1}{2}\int_v \rho_{DM}\phi_{DE} \,dv \nonumber \\
    && + \frac{1}{2}\int_v \rho_{DE}\phi_{DM} \,dv+ \frac{1}{2}\int_v \rho_{DE}\phi_{DE}\,dv\,, \nonumber\\
\label{VT1}
\end{eqnarray}
Here $\phi_{DM}$ and $\phi_{DE}$ are the gravitational potentials of dark matter and the dark energy components and $\rho_{DM}$ and $\rho_{DE}$ are overdensities of the dark matter and the dark energy components. The four distinguishable scenarios arising from the non-zero values of these integrations can be categorized as follows:

\begin{itemize}
  \item Isolated Sub-Universe (No Dark Energy Clustering):
    In this scenario, only the first integration in Eq.~(\ref{VT1}) is non-zero. Spherical over-densities of dark matter behave like an isolated sub-universe and virialize at a certain radius, and this scenario aligns precisely with what the top-hat collapse model describes \cite{Gunn}.

  \item Homogeneous Dark Energy Model:
This scenario arises when the first two integrations in Eq.~(\ref{VT1}) contribute to the total gravitational potential. The non-zero values of the second integration in Eq.~(\ref{VT1}) imply a Non-negligible effect of dark energy on the virialization process of spherically symmetric over-dense regions of dark matter. However, in this scenario, dark energy cannot cluster and virialize with dark matter; the dark energy density inside the over-dense region remains similar to the external dark energy density \cite{Lahav:1991wc, Steinh, Shapiro, Horellou:2005qc}.

  \item Clustered Dark Energy Scenario:
In the third scenario, dark energy doesn't undergo virialization with dark matter, yet it can cluster within over-dense regions. Here, it is assumed that since the beginning of the matter-dominated era, dark energy synchronously follows the motion of dark matter on both the Hubble scale and the galaxy cluster scale. This scenario is referred to as the clustered dark energy scenario \cite{Basilakos:2003bi, Maor:2005hq, Basilakos:2006us, Basilakos:2009mz, Chang:2017vhs, Dey2}.

  \item Dark Energy Clustering and Virialization:
In this scenario, dark energy can cluster and virialize with dark matter inside the spherical over-dense regions \cite{Maor:2005hq}.
\end{itemize}
In \cite{Lahav:1991wc} and \cite{Steinh}, the authors studied a cosmological scenario where dark energy exhibits homogeneity, meaning that internal and external dark energy densities are identical. In \cite{Lahav:1991wc}, the focus was on investigating the impact of the cosmological constant on the virialization of spherical over-densities. Whereas, L. Wang {\it et al.} in \cite{Steinh} considered the homogeneous quintessence dark energy model. As mentioned earlier, in scenarios with homogeneous dark energy, dark energy doesn't cluster or virialize within the spherical over-densities of dark matter. However, the virialization process of these over-densities is altered due to the presence of non-zero energy density and negative pressure of dark energy, leading to distinct values of $\eta$ which is, as mentioned previously, the ratio between the $R_{vir}$ and $R_{max}$. It can be shown that the $\eta$ is always less than $0.5$ when $\Lambda$ dark energy is considered \cite{Lahav:1991wc, Shapiro, Saha1}. In the homogeneous dark energy model, a significant challenge arises because, following the virialization of over-dense regions, the density of dark energy within the virialized region continues to change with the continuous expansion of the background universe. This issue is thoroughly examined in \cite{Maor:2005hq}. However, this problem does not appear for the $\Lambda$ dark energy scenario, since the energy density of the dark energy remains constant throughout the evolution.
In order to resolve the problem with homogeneous dark energy, clustered dark energy models are introduced, where, at the scale of galaxy clusters, dark energy can cluster and virialize within over-dense regions.
For this scenario, one can verify that $\eta$ is always greater than $0.5$. Whereas, in the case where dark energy can cluster but cannot virialize inside the over-dense region, the virialized radius of the spherical over-dense region becomes smaller than half of the turnaround radius (i.e., $\eta < 0.5$) \cite{Maor:2005hq, Saha1}. In \cite{Saha1}, we show that the above-mentioned behavior of $\eta$ is true for both the quintessence and phantom-like dark energy. 

For the homogeneous dark energy scenario, the problem that we discussed above is addressed in most of the literature devoted to this topic, by not considering the formal general relativistic approach but instead using a pure phenomenological method. In this method, the problem is approached non-relativistically. It involves using an FLRW metric with a positive spatial curvature constant and subsequently formulating the Friedmann equations. The first Friedmann equation, which incorporates the square of the first derivative of the local scale factor, poses challenges, especially as it necessitates estimating all the known energy sources within the spherical patch. Given that energy may not be conserved, this equation becomes redundant. The majority of prior research in this domain relies predominantly on the other Friedmann equation, which encompasses the second derivative of the scale factor. This equation is treated as a second-order ordinary differential equation in time and is solved with suitable initial conditions.

In recent work \cite{Saha1}, we addressed this issue of the homogeneous dark energy model by employing a more relativistic method. In that paper, we studied the evolution of over-dense regions of dark matter in the presence of a minimally coupled scalar field representing homogeneous dark energy. We tackled the problem in the homogeneous dark energy model by matching the internal closed FLRW spacetime with an external generalized Vaidya space-time resulting in a leaking of scalar field through the boundary of the over-dense region. The non-zero flux of the scalar field through the boundary of the over-dense region shows how the dark energy retains its homogeneous nature throughout the evolution. Therefore, in the regime of general relativity, using our model, we showed how homogeneous dark energy influences the virialization process of the dark matter. However, it should be noted that our method is relativistic up to the virialization. We employed the Newtonian virialization technique to investigate the virialized end states of the over-dense regions. 

In this paper, we investigate the evolution of the over-dense region of dark matter in the presence of a non-minimally coupled scalar field. We adopt a spacetime structure similar to the one examined in \cite{Saha1}. This choice is crucial as the specified spacetime structure plays an important role in preserving the homogeneous behavior, if indeed it exists, of dark energy modeled by a non-minimally coupled scalar field. Therefore, in the present paper, we consider an external generalized Vaidya spacetime which is smoothly matched at the boundary of the internal closed FLRW spacetime. It is important to emphasize that our focus here is solely on the algebraic coupling between the scalar field and matter. By algebraic coupling, we mean that the interaction Lagrangian does not depend on derivatives of the scalar field \cite{Tamanini1, Tamanini2, Tamanini3}. Utilizing algebraic non-minimal coupling, we explore the evolution of over-dense regions of dark matter in the presence of quintessence and phantom-like dark energy.
The main motivation behind examining the non-minimal coupling between dark matter and dark energy is to gain insights into how this coupling can impact the virialized structures of dark matter on a certain cosmological scale where the influence of dark energy cannot be ignored. As previously mentioned, our earlier study focused on minimal coupling, and the results diverged significantly from those obtained with the top-hat model. These disparities prompt us to investigate the same scenario with non-minimal coupling. Similar to our prior study \cite{Saha1}, we observe that, for some suitable small values of parameters, the dark energy component remains predominantly unclustered and homogeneous. However, our findings also reveal that an increment of the non-minimal coupling between dark matter and dark energy leads to the clustering of dark energy within the over-dense region of dark matter. This clustering arises due to the specific nature of the non-minimal coupling considered in our study.
It can be shown that the energy density of matter remains unaffected by the non-minimal interaction, staying proportional to $\frac{1}{a(t)^3}$, where $a(t)$ is the scale factor. On the other hand, the Klein-Gordon equation of the scalar field undergoes modification with an additional interaction term. Consequently, an increase in the coupling strength compels the dark energy to deviate from its homogeneous state and cluster with the dark matter inside the over-dense region. In essence, we can describe this phenomenon as the dark matter pulling the dark energy inward as the coupling strength increases. 

As the scenario with minimal coupling, which we investigated earlier, is a subset of the current scenario, similar to the previous case, closed spherical regions do not undergo collapse for certain parameter ranges; instead, they exhibit eternal expansion within the spatially flat FLRW spacetime. These expanding regions act as voids, with decreasing matter density. Like our previous study, here our approach is relativistic up to the virialization. We consider the Newtonian virialization technique to stabilize the collapsing over-dense region of dark matter. The quest for a comprehensive understanding of the general relativistic counterpart to the Newtonian virialization process remains a challenging problem. Dey et al. \cite{Dey2} present a dynamic solution within the framework of general relativity, illustrating a gravitational collapse leading to an equilibrium state. However, they do not assert that this equilibrium state directly corresponds to the Newtonian virialization state. Recently, in \cite{Dey:2023laa} the authors derived the possible form of the scalar field potential which can lead to an end equilibrium state of the gravitational collapse of the scalar field, using the general relativistic equilibrium conditions mentioned in \cite{Dey2}. Meyer et al. \cite{Meyer1} introduce a general relativistic virial theorem based on the Tolman-Oppenheimer-Volkoff (TOV) solution for perfect-fluid spheres in the Einstein-de Sitter and $\Lambda$CDM cosmologies. However, they do not clarify how a collapsing matter cloud reaches the virialization state. In another work \cite{Friedman1}, Friedman and Stergioulas introduce a virial theorem definition in stationary spacetimes, explored in Section (3.3) of their monograph. A focused exploration into deriving the relativistic virial condition for dynamic spacetimes, building upon the introduced condition in the monograph, could offer valuable insights.

The work in this paper is organized in the following way. In section \ref{sec2}, we elaborately discuss the non-minimal coupling between the matter and the scalar field, where we review the basic foundation of the works done in \cite{Tamanini1, Tamanini2, Tamanini3}. In Section \ref{sec3}, we discuss the spacetime structure considered in this paper and explore the impact of non-minimal coupling on the evolution of the over-dense region of dark matter by solving a differential equation derived from the Friedmann equations. At last, in that section, we discuss the results and their possible physical interpretation. Section \ref{sec4} gives a summary of the work presented in this paper. Throughout the paper, we use a system of units in which the velocity of light and the universal gravitational constant (multiplied by $8\pi$), are both set equal to unity.

\section{Non-minimal coupling of matter with scalar field}\label{sec2}
In this section, we briefly discuss the non-minimal coupling of matter with a scalar field which is worked out elaborately in \cite{Tamanini1, Tamanini2, Tamanini3}. The action we will consider is,

\begin{eqnarray}
\mathcal{S}=\int d^4x~\left(\mathcal{L}_{\text{grav}}+\mathcal{L}_{m}+\mathcal{L}_{\phi}+\mathcal{L}_{\text{int}}\right)\,\, ,
\end{eqnarray}
where the gravitational sector is given by the standard Einstein–Hilbert Lagrangian:
\begin{eqnarray}
\mathcal{L}_{\text{grav}}=\frac{\sqrt[•]{-g}R}{2}\,\, ,
\end{eqnarray} 
where $g$ is the determinant of the metric tensor $g_{\mu\nu}$ and $R$ is the Ricci scalar.
Within Brown’s framework, the Lagrangian for the relativistic fluid can be written as
\begin{eqnarray}\label{1}
\mathcal{L}_{m}=-\sqrt[]{-g}\rho_{m}(n,s)+J^{\mu}\left(\varphi_{,\mu}+s\theta_{,\mu}+\beta_{A}\alpha^{A}_{,\mu}\right)\,\, ,
\end{eqnarray}
 where $\rho_{m}$ is the energy density of the matter. We assume $\rho_{m}(n,s)$ to be prescribed as a function of n, the particle number density, and s, the entropy density per particle. $\varphi$, $\theta$, and $\beta_{A}$ are all Lagrange multipliers with A taking the values 1,2,3, and $\alpha^{A}$ are the Lagrangian coordinates of the fluid. The vector density or the current density of particle number $J^{\mu}$ is related to n as
$$J^{\mu}=\sqrt[]{-g}nU^{\mu},~ |J|=\sqrt[]{-g_{\mu\nu} J^\mu J^\nu},~ n=\frac{|J|}{\sqrt[]{-g}}\, ,$$
where $U^{\mu}$ is the timelike 4-velocity of matter satisfying $U_{\mu}U^{\mu}=-1$.
The scalar field Lagrangian is given by
\begin{eqnarray}\label{2}
\mathcal{L}_{\phi}=-\sqrt[]{-g}~\left[\frac{1}{2}\epsilon\partial_{\mu}\phi \partial^{\mu}\phi+V(\phi)\right]\, ,
\end{eqnarray}
where $\epsilon=1,-1$ are for quintessence and phantom-like scalar field, respectively and $V(\phi)$ is the potential of the scalar field $\phi$. Lastly, the Lagrangian for the interacting sector is
\begin{eqnarray}\label{3}
\mathcal{L}_{\text{int}}=-\sqrt[]{-g}f(n,s,\phi)\, ,
\end{eqnarray}
where $f(n,s,\phi)$ is an arbitrary function of $n,~s$ and $\phi$.
Now as we know, the total energy-momentum tensor can be written as:
\begin{eqnarray}\label{4}
T_{\mu\nu}=\frac{-2}{\sqrt[]{-g}}\frac{\delta \mathcal{L}}{\delta g^{\mu\nu}}.
\end{eqnarray}
Therefore, the energy-momentum tensors for the scalar field, matter, and interaction part are
\begin{eqnarray}\label{5}
T^{(\phi)}_{\mu\nu}&=&\epsilon\partial_{\mu}\phi \partial_{\nu}\phi-g_{\mu\nu}\left[\frac{1}{2}\epsilon\partial_{\mu}\phi \partial^{\mu}\phi+V(\phi)\right]\, ,\\
\label{6}
T^{(m)}_{\mu\nu}&=&p_{m}g_{\mu\nu}+(\rho_{m}+p_{m})U_{\mu}U_{\nu}\, ,\\
\label{7}
T^{(\text{int})}_{\mu\nu}&=&p_{\text{int}}g_{\mu\nu}+(\rho_{\text{int}}+p_{\text{int}})U_{\mu}U_{\nu},
\end{eqnarray}
where the pressure in the matter sector $p_{m}$ can be written as
$$p_{m}=n\pdv{\rho_{m}}{n}-\rho_{m},$$
$$\rho_{\text{int}}=f(n,s,\phi)$$ and $$p_{\text{int}}=n\pdv{f(n,s,\phi)}{n}-f(n,s,\phi).$$
Now, the total energy-momentum tensor for the physical system can be written as
$$T^{(\text{t})}_{\mu\nu}= T^{(m)}_{\mu\nu}+T^{(\phi)}_{\mu\nu}+T^{(\text{int})}_{\mu\nu}$$ and the Einstein equation gives us:
\begin{eqnarray}\label{11}
G_{\mu\nu}=T^{(\text{t})}_{\mu\nu}.
\end{eqnarray}
Variation of the Lagrange multipliers in the total Lagrangian gives
\begin{eqnarray}
J^\mu: U_{\mu}(\mu_{\text{int}}+\mu)+\varphi_{,\mu}+s\theta_{,\mu}+\beta_{A}\alpha^{A}_{,\mu}=0,
\end{eqnarray}
\begin{eqnarray}
s: -\left[\frac{\partial\rho}{\partial s}+\frac{\partial f}{\partial s}\right]+nU^\mu \theta_{,\mu}=0,
\end{eqnarray}
\begin{eqnarray}\label{a}
\phi:  J^\mu_{,\mu}=0,
\end{eqnarray}
\begin{eqnarray}\label{b}
\theta: \left(sJ^\mu\right),_{\mu}=0,
\end{eqnarray}
\begin{eqnarray}
\beta_{A}: J^{\mu}\alpha^{A}_{,\mu}=0,
\end{eqnarray}
\begin{eqnarray}
\alpha^{A}: (\beta_{A}J^{\mu})_{,\mu}=0,
\end{eqnarray}
where $\mu$ is the chemical potential given by $\mu=\frac{\partial\rho}{\partial n}$ and $\mu_{int}= \frac{\partial f}{\partial n}$. Here $\mu_{int}$ is a new variable defined for our case and it is not the standard chemical potential as $\mu = (\rho+p)/n$ \cite{Tamanini1, Tamanini2, Tamanini3}.

Eqns. (\ref{a}) and (\ref{b}) stand for the particle number conservation constraint
and the entropy exchange constraint, respectively. Both of these can be written as
\begin{eqnarray}\label{numberandentropyequ}
\nabla_{\mu}(nU^\mu)=0 &~\text{and}~& \nabla_{\mu}(snU^\mu)=0.
\end{eqnarray}
Now, the modified Klein-Gordon equation is,
\begin{eqnarray}\label{KG1}
\square\phi-\pdv{V}{\phi}-\pdv{f}{\phi}=0.
\end{eqnarray}
\section{Spacetime structure and the governing Cosmological equations}
\label{sec3}
\subsection{Spacetime configuration} In this paper, as we mentioned before, in order to model the dynamics of the over-dense region of dark matter in the presence of non-minimally coupled dark energy, we use closed FLRW spacetime:
\begin{eqnarray} \label{FLRW1}
ds^{2}=-dt^{2}+\frac{a^{2}(t)}{1-kr^{2}}dr^{2}+r^{2}a^{2}(t)(d\theta^{2}+\sin^{2}\theta d\Phi^{2})\,\,,\nonumber\\ 
\end{eqnarray}
where the constant $k$ can be $0,\pm1$ and $a(t)$ is the scale factor of the over-dense region. A value of $k=0$ signifies a flat spatial component, while negative and positive values indicate an open or closed spatial section, respectively. We consider closed FLRW metric to model the over-dense region of dark matter since the dynamics of a flat universe are always monotonic. A flat universe either expands or collapses depending upon the initial values of $\dot{a}(t)$ i.e., the flat universe cannot have a turnaround scenario if we do not include bounces. 

Like the minimal coupling scenario we studied earlier \cite{Saha1}, here also we want to generalize the top-hat collapse model in the presence of non-minimally coupled dark energy, and therefore, we choose closed FLRW spacetime to model the over-dense region. At the initiation of the gravitational collapse ($t=0$), $a(t)$ can assume any positive definite value, which can always be rescaled to one. Therefore, we set $a(t=0) = 1$. To account for the presence of dark matter and dark energy in the ever-expanding background of the over-dense regions, we model the background using the  flat FLRW spacetime:
\begin{eqnarray} \label{flatFLRW}
ds^{2}=-dt^{2}+\bar{a}^{2}(t)dr^{2}+r^{2}\bar{a}^{2}(t)(d\theta^{2}+\sin^{2}\theta d\Phi^{2})\,\,,\nonumber\\ 
\end{eqnarray}
where the scale factor of the background is denoted by $\bar{a}(t)$. Henceforth, any parameter with an over-bar denotes its association with the background. As done in our previous work, we utilize an external generalized Vaidya space-time to depict the matter flux through the boundary of the over-dense region within its immediate vicinity. Importantly, Vaidya space-time is not regarded as a background space-time; rather, the background at the Hubble scale is modeled using flat FLRW space-time. Vaidya space-time is exclusively employed to depict the localized dynamics of matter around the boundary of the over-dense regions.
Therefore, in our model, at a timelike hypersurface $\Sigma = r -r_b =0,~ \forall r_b < 1$,  the internal closed FLRW spacetime smoothly matches the external generalized Vaidya spacetime:
\begin{eqnarray}
dS^2_{-}&=&-dt^2+a^2(t)\left(\frac{dr^2}{1-r^2}+r^2d\Omega^2\right)\,\,\nonumber\\
&=& -dt^2 + a^2(t) d\Psi^2 +a^2(t)\sin^2\Psi d\Omega^2\,\, ,\\
\nonumber\\
dS^2_{+}&=& -\left(1-\frac{2M(r_v , v)}{r_v}\right)dv^2 - 2dv dr_v + r_v^2 d\Omega^2\,\, .\nonumber\\
\end{eqnarray}
Here, we define the co-moving radius as $r=\sin\Psi$ and $r_v$ along with $v$ represent the coordinates associated with the generalized Vaidya spacetime. For the smooth matching at $\Sigma$, as we know, we need to match the induced metric ($h_{ab}$) and the extrinsic curvature ($K_{ab}$) on the $\Sigma$ from both sides. From the induced metric matching we get:
\begin{eqnarray}
\left(\dot{v}^2-\frac{2M(r_v , v)}{r_v}\dot{v}^2+2\dot{v}\dot{r}_v\right) &=& 1\,\, ,\\
r_v &=& a(t)\sin\Psi_b\,\, ,
\end{eqnarray}
and the matching of the extrinsic curvature gives us \cite{Saha1}:
\begin{eqnarray}
\cos\Psi_b &=& \frac{1-\frac{2M}{r_v}+\frac{dr_v}{dv}}{\sqrt{1-\frac{2M}{r_v}+2\frac{dr_v}{dv}}}\, ,\\
\label{Massderivative}
M(r_v,v)_{,r_v}&=&\frac{F}{2\sin\psi_b a(t)}+\sin^2\psi_b a\ddot{a}\,\, .
\end{eqnarray}
Here, $F$ denotes the Misner-Sharp mass of the internal collapsing spacetime, and it must satisfy the following condition at the boundary:
\begin{equation}
F(t,\sin\psi_b)=2M(r_v,v)\,\,.
\label{FM1}
\end{equation}
The matter flux at the boundary depends on the scale factor and the Misner-Sharp mass ($F$) of the collapsing spacetime, as indicated in Eq.~(\ref{Massderivative}). In this case, the Misner-Sharp mass $F$ of the internal spacetime is a time-dependent function only since the internal spacetime is spatially homogeneous. As we know, the internal pressure $p = - \frac{\dot{F}}{\dot{R}R^2}$, and therefore, the existence of non-zero pressure at the boundary signifies a non-zero matter flux through it. This is why we incorporate the use of the generalized Vaidya spacetime in the immediate vicinity of the internal two-fluid system. The negative pressure at the boundary implies an inward matter flux for an expanding scenario and an outward matter flux for a collapsing scenario. It can be seen in the next subsection that the non-minimal coupling we consider here does not have any impact on the evolution of the matter part, whereas it modifies the Klein-Gordon equation of the scalar field. Therefore, dust-like matter evolves like $\frac{1}{a^3(t)}$, and there is zero-flux of the matter at the boundary of the over-dense region. Only the scalar field leaks out of the boundary since the pressure at the boundary is negative. This result is similar to the results we demonstrated in our previous work with minimal coupling. However, it is also shown in the next subsection that the non-minimal coupling indeed causes dark energy to deviate from its homogeneous nature. An increment of the strength of the non-minimal coupling results in a lesser outward flux of the scalar field, and consequently, $\frac{\rho_{\phi}}{\bar{\rho}_{\phi}}$ becomes greater than one. Therefore, the non-minimal coupling forces the dark energy to cluster inside the over-dense region. 
\subsection{Governing cosmological equations of the non-minimally coupled matter and scalar field}
From (\ref{numberandentropyequ}), one can obtain the conservation equation for the number density and entropy as
\begin{eqnarray}
\dot{n}+3Hn=0 &~\text{and}~& \dot{s}=0\, .
\end{eqnarray}
Now, from equation (\ref{6}), (\ref{7}) and (\ref{5}), we get $$T^{\mu (m)}_{\nu}=diag(-\rho_{m},p_{m},p_{m},p_{m})\, ,$$ $$T^{\mu (\text{int})}_{\nu}=diag(-\rho_{\text{int}},p_{\text{int}},p_{\text{int}},p_{\text{int}})$$
and $$T^{0 (\phi)}_{0}=-\left[\frac{1}{2}\epsilon\dot\phi^2+V(\phi)\right],~ T^{i (\phi)}_{i}=\frac{1}{2}\epsilon\dot\phi^2-V(\phi).$$
Substituting the above expressions for the energy-momentum tensors of matter, scalar field, and the interaction component into the Friedman equations associated with the internal metric, we obtain:
\begin{eqnarray}\label{one}
\frac{3\dot{a}^2}{a^2}+\frac{3k}{a^2}=\left(\rho_{m}+\frac{1}{2}\epsilon\dot\phi^2+V(\phi)+\rho_{\text{int}}\right)\, ,
\end{eqnarray}
\begin{eqnarray}\label{two}
\frac{2\ddot{a}}{a}+\frac{\dot{a}^2}{a^2}+\frac{k}{a^2}=-\left(p_{m}+\frac{1}{2}\epsilon\dot\phi^2-V(\phi)+p_{\text{int}}\right)\, .
\end{eqnarray}
Now, from the conservation of the total energy-momentum tensor, we can write:
\begin{eqnarray}
\nabla_{\mu}T^{\mu\nu}_{(\text{t})}=\nabla_{\mu}T^{\mu\nu}_{(m)}+\nabla_{\mu}T^{\mu\nu}_{(\phi)}+\nabla_{\mu}T^{\mu\nu}_{(\text{int})} = 0,
\end{eqnarray}
which implies:
\begin{eqnarray}
    \dot{\rho}_{m} + 3\frac{\dot{a}}{a}(\rho_{m}+p_{m})~+~\dot{\phi}\left(\ddot{\phi}+3H\dot{\phi}+\frac{\partial V}{\partial\phi}+\frac{\partial \rho_{\text{int}}}{\partial\phi}\right) = 0.\nonumber\\
\end{eqnarray}
Therefore, the modified Klein-Gordon equation (Eq.~(\ref{KG1})) becomes:
\begin{eqnarray}\label{Klein1}
\ddot{\phi}+3H\dot{\phi}+\frac{\partial V}{\partial\phi}+\frac{\partial \rho_{\text{int}}}{\partial\phi}=0,
\end{eqnarray}
and the matter equations of motion is
\begin{eqnarray}\label{mEq}
\dot{\rho}_{m} + 3\frac{\dot{a}}{a}(\rho_{m}+p_{m})=0.
\end{eqnarray}
From above Eqs.~(\ref{Klein1}, \ref{mEq}), it can be observed that the non-minimal coupling alters the Klein-Gordon equation, while the equation of motion for matter energy density remains similar to that in the minimal coupling scenario.
Now, if we take the matter as dust then pressure $p_{m}=0$ and from (\ref{one}) and (\ref{two}) we will get, 
\begin{eqnarray}\label{three}
\frac{3\dot{a}^2}{a^2}+\frac{3k}{a^2}=\left(\rho_{m}+\frac{1}{2}\epsilon\dot\phi^2+V(\phi)+\rho_{\text{int}}\right)\, ,
\end{eqnarray}
\begin{eqnarray}\label{four}
\frac{2\ddot{a}}{a}+\frac{\dot{a}^2}{a^2}+\frac{k}{a^2}=-\left(\frac{1}{2}\epsilon\dot\phi^2-V(\phi)+p_{\text{int}}\right)\, ,
\end{eqnarray}
and with $p_m = 0$, Eq.~(\ref{mEq}) implies $\rho_m = \frac{\rho_{m_{0}}}{a^3(t)}$, where $\rho_{m_{0}}$ is the initial density of matter in the over-dense region. 
Using 
$\rho_{\phi}=\frac{1}{2}\epsilon\dot\phi^2+V(\phi)$ in (\ref{three}), we get
\begin{eqnarray}\label{five}
\dot{a}=\pm\sqrt[]{\frac{(\rho_{m}+\rho_{\phi}+\rho_{\text{int}})a^{2}}{3}-k}\, .
\end{eqnarray}
Differentiating (\ref{five}) with respect to co-moving time (t) we get
\begin{eqnarray}\label{six}
\ddot{a}=\frac{a}{3}\left[\rho_{m}+\rho_{\phi}+\rho_{\text{int}}+\frac{a}{2}(\rho_{m,a}+\rho_{\phi,a}+\rho_{\text{int},a})\right]\, ,\nonumber\\
\end{eqnarray}
where $\rho_{\phi,a}$, $\rho_{m,a}$ and $\rho_{int,a}$ are derivatives of the scalar field energy density, the fluid energy density and the interaction part energy density respectively, with respect to the scale factor a.
Similarly, using $p_{\phi}=\frac{1}{2}\epsilon\dot\phi^2-V(\phi)$, we obtain
\begin{eqnarray}\label{seven}
\rho_{\phi}+p_{\phi}=\epsilon\phi_{,a}^{2}\dot{a}^2 &~\text{and}~& p_{\phi}=\rho_{\phi}-2V(\phi)\, .
\end{eqnarray}
Substituting the expression of $\dot{a}$ from Eq.~(\ref{five}) in the Eq.~(\ref{seven}) we get
\begin{equation}\label{nine}
\rho_{\phi}\left(1-\frac{\epsilon\phi_{,a}^{2}a^{2}}{3}\right)-\left(\rho_{m}+\rho_{int}\right)\frac{\epsilon\phi_{,a}^{2}a^{2}}{3}+p_{\phi}+k\epsilon\phi_{,a}^{2}=0\, .
\end{equation}
Now, putting the value of $p_{\phi}$ from Eq.~(\ref{seven}) in Eq.~(\ref{nine}) we get
\begin{equation}\label{ten}
\rho_{\phi}=\frac{\frac{(\rho_{m}+\rho_{int})\epsilon\phi_{,a}^{2}a^{2}}{6}+V(\phi)-\frac{k\epsilon\phi_{,a}^{2}}{2}}{\left(1-\frac{\epsilon\phi_{,a}^{2}a^{2}}{6}\right)}\, .
\end{equation}
\begin{figure*}
\centering
\subfigure[Region plot with $\alpha=3$]
{\includegraphics[width=82mm,height=50mm]{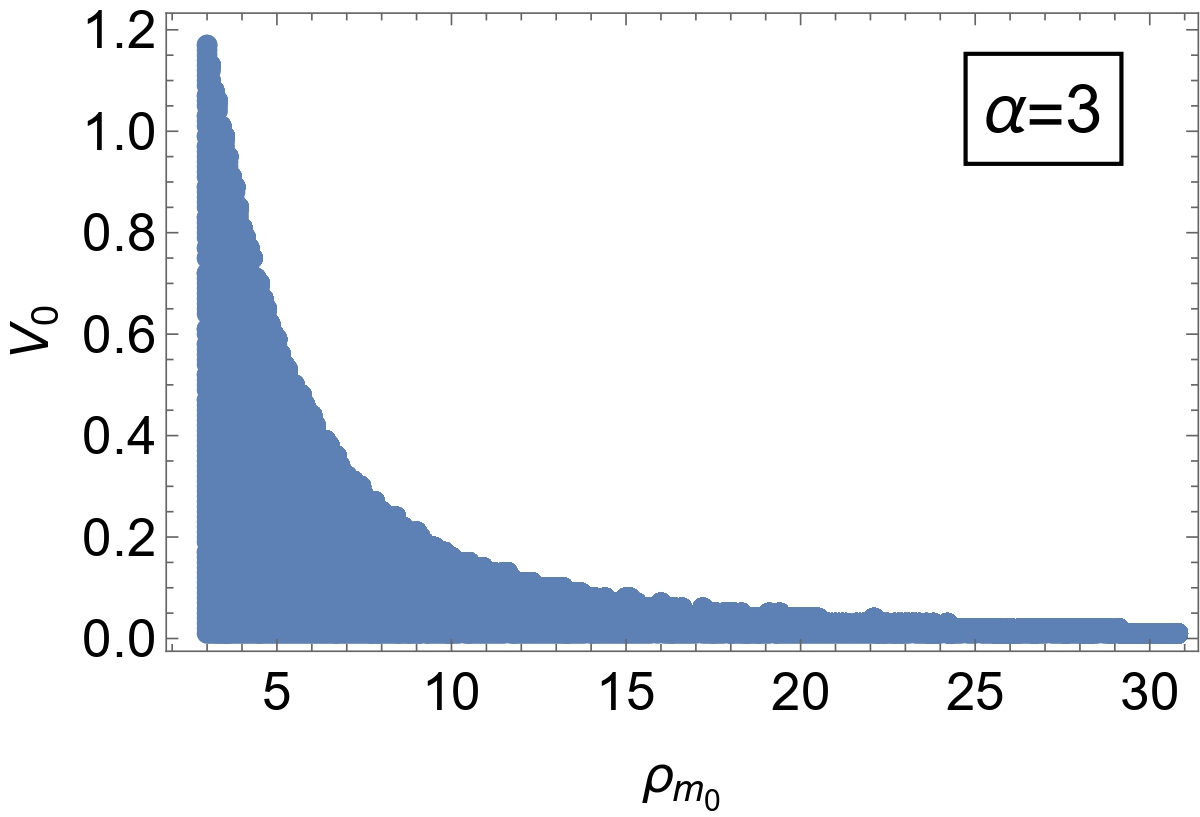}\label{Fig1}}
\subfigure[Region plot with $\alpha=6$]
{\includegraphics[width=82mm,height=50mm]{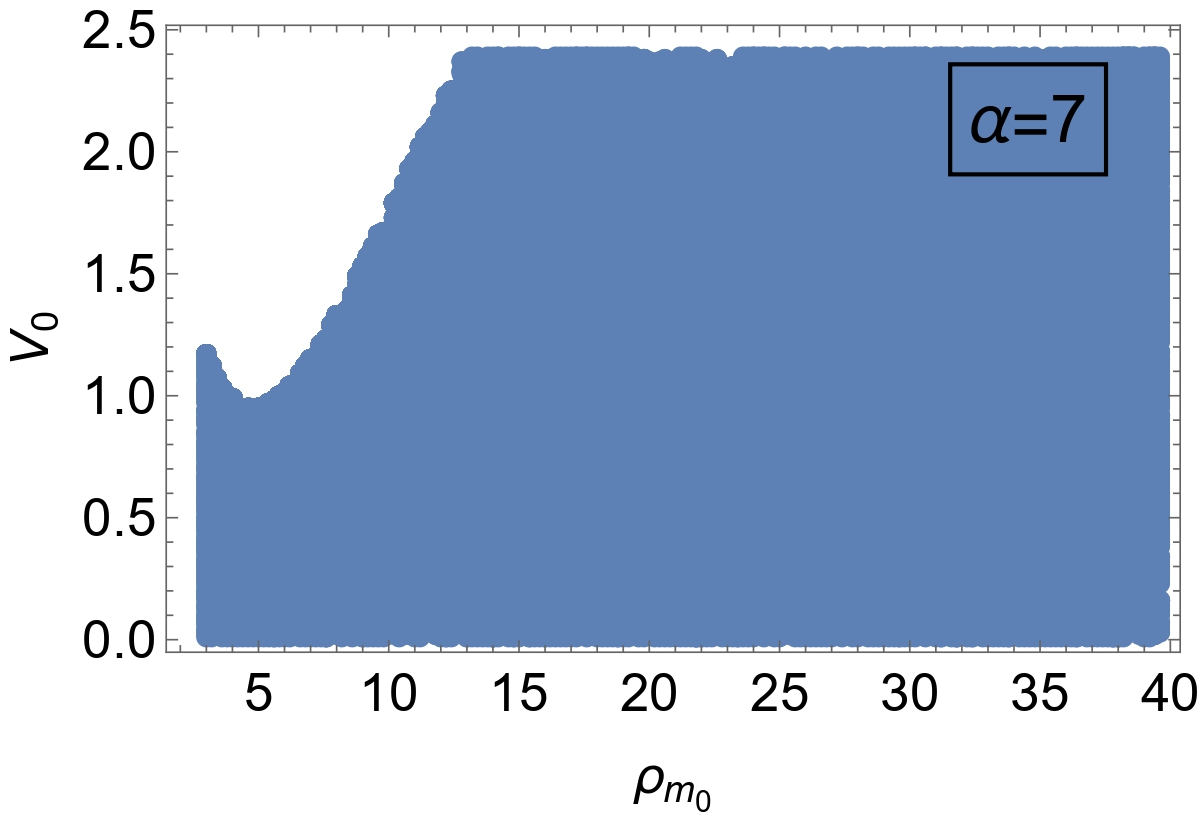}\label{Fig2}}
\caption{Figure shows variation of Region plot of $V_{0}$ vs $\rho_{m_{0}}$ for increasing values of $\alpha$ for Quintessence field.}
\label{RegionplotforvariationofQalpha}
\end{figure*}
\begin{figure*}
\centering
\subfigure[Region plot with $\alpha=3$]
{\includegraphics[width=82mm,height=50mm]{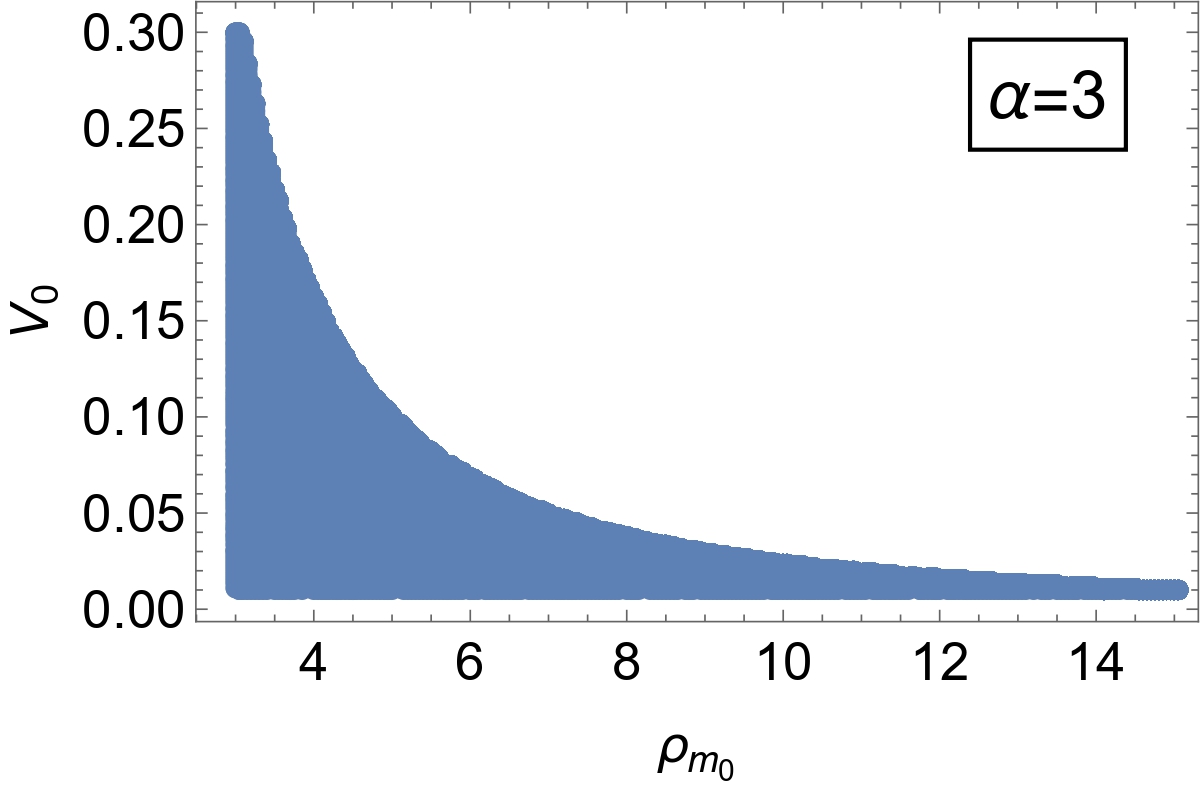}\label{Fig3}}
\subfigure[Region plot with $\alpha=7$]
{\includegraphics[width=82mm,height=50mm]{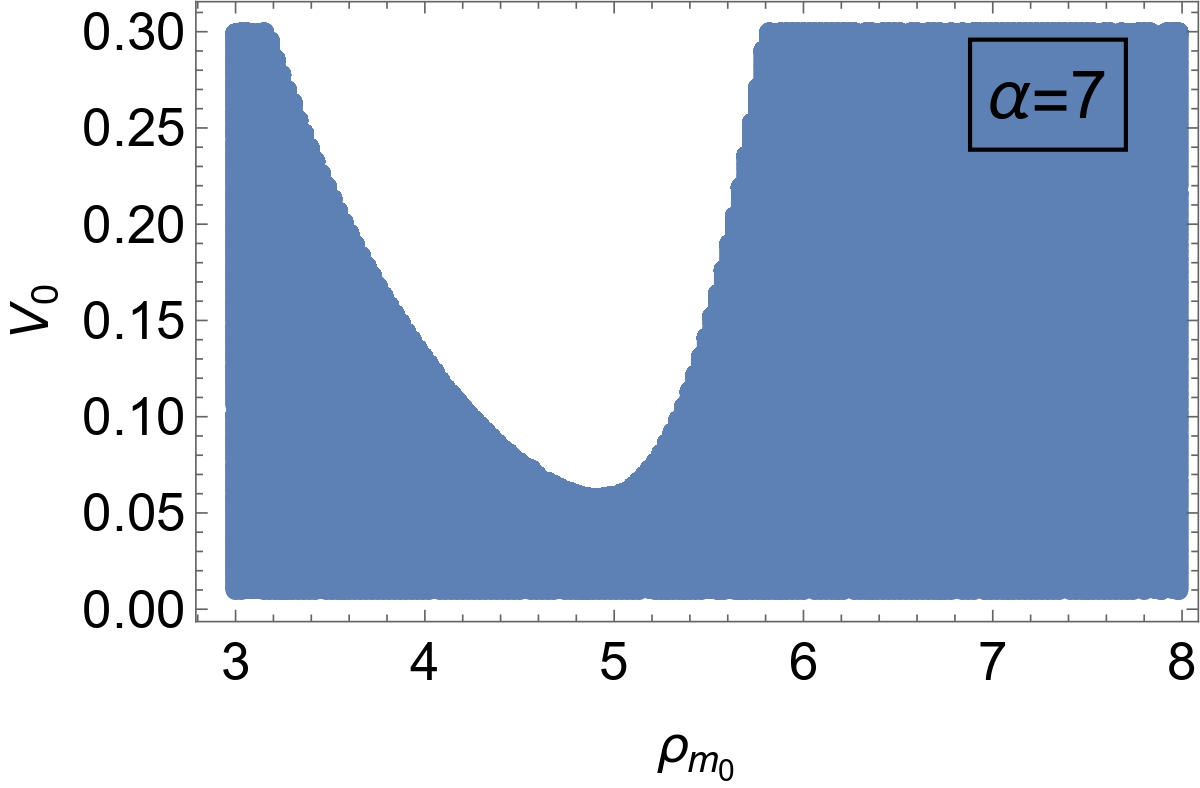}\label{Fig4}}
\hspace{0.2cm}
\caption{Figure shows variation of Region plot of $V_{0}$ vs $\rho_{m_{0}}$ for increasing values of $\alpha$ for Phantom field}
\label{Regionplotforvariationofphalpha}
\end{figure*}

Now using (\ref{four}), (\ref{five}), (\ref{six}), (\ref{seven}) we get
\begin{eqnarray}\label{tweleve}
\rho_{\phi ,a}&=&-\left[\left(\frac{\rho_{m}}{a}+\frac{\rho_{\text{int}}}{a}\right)\left(3+\epsilon\phi_{,a}^{2}a^{2}\right)+\rho_{\phi}\epsilon\phi_{,a}^{2}a+\frac{3p_{int}}{a}-\frac{3k\epsilon\phi_{,a}^2}{a}\right]\nonumber\\
&-&\rho_{m,a}-\rho_{\text{int},a}\, .
\end{eqnarray}
Now, differentiating Eq. (\ref{ten}) with respect to $a$ and using equation Eq. (\ref{tweleve}) we obtain the following second order
differential equation
\begin{widetext}
\begin{eqnarray}  \label{eleven}
-\left[\frac{(\rho_{m}+\rho_{\text{int}})(3+\epsilon\phi_{,a}^{2}a^{2})+\rho_{\phi}\epsilon\phi_{,a}^{2}a^{2}+3p_{\text{int}}-3k\epsilon\phi_{,a}^{2}}{a}\right]-\rho_{m,a}-\rho_{\text{int},a}&=&\frac{1}{3(1-\frac{\epsilon\phi_{,a}^{2}a^{2}}{6})^{2}}\{3V_{,\phi}\phi_{,a}+\frac{(\rho_{m,a}+\rho_{\text{int},a})\epsilon\phi_{,a}^{2}a^{2}}{2}\nonumber\\&-&\frac{(\rho_{m,a}+\rho_{\text{int},a})\epsilon^{2}\phi_{,a}^{4}a^{4}}{12}+(\rho_{m}+\rho_{\text{int}})\epsilon \phi_{,a}^{2}a\nonumber\\&+&(\rho_{m}+\rho_{\text{int}})\epsilon\phi_{,a}\phi_{,aa}a^{2}-\frac{\epsilon V_{,\phi}\phi_{,a}^{3}a^{2}}{2}+\epsilon V(\phi)a\phi_{,a}^{2}\nonumber\\&+&\epsilon V(\phi)a^{2}\phi_{,a}\phi_{,aa}-3k\epsilon\phi_{,a}\phi_{,aa}-\frac{k\epsilon^{2}\phi_{,a}^{4}a}{2}\}\,\, ,\nonumber\\
\end{eqnarray}
\end{widetext}

\subsection{Construction of the model}

We aim to characterize our system using an autonomous system of differential equations involving three variables, ensuring its self-containment as done in \cite{Tamanini1}. The autonomous equations take the form:
$$
x' = f_{1}(x, y, z), \quad y' = f_{2}(x, y, z), \quad z' = f_{3}(x, y, z),
$$
where \(x\), \(y\), and \(z\) represent three dynamic variables of the system, and the prime notation denotes differentiation with respect to the time parameter. The functions \(f_{i}(x, y, z)\) for \(i = 1, 2, 3\) do not explicitly depend on the time parameter; they solely rely on \(x\), \(y\), and \(z\).

The background evolution modeled by flat FLRW spacetime can be parameterized in terms of phase-space variables \(\sigma\), \(X\), \(Y\), and \(Z\), defined as follows:

$$
X = \frac{\dot{\phi}}{\sqrt{6}H}, \quad Y = \frac{\sqrt{V(\phi)}}{\sqrt{3}H}, \quad Z = \frac{\rho_{\text{int}}}{3H^{2}}, \quad \sigma = \frac{\sqrt{\rho_{m}}}{\sqrt{3}H}.
$$
These variables allow the background Friedmann equation to be expressed as:
$$
1 = \sigma^{2} + X^{2} + Y^{2} + Z,
$$
serving as a constraint equation permitting the replacement of \(\sigma\) with other variables. With these variables, the cosmological field equations can be written in terms of \(X'\), \(Y'\), \(Z'\), \(X\), \(Y\), \(Z\), \(A\), and \(B\), where prime denotes the derivative with respect to \(Hdt\). Here,
$$
A = \frac{p_{\text{int}}}{2H^{2}}, \quad B = \frac{1}{\sqrt{6}H^{2}}\frac{\partial\rho_{\text{int}}}{\partial\phi}.
$$

In selecting our model, we define functions \(\rho_{\text{int}}\) and \(p_{\text{int}}\) such that \(A\) and \(B\) become functions of \(X\), \(Y\), and \(Z\) only. This ensures the system is closed at both the background and perturbation levels, and consequently, we can write the following expression of $\rho_{\text{int}}$ and $p_{\text{int}}$ \cite{Tamanini1}:
\begin{eqnarray}
    \rho_{\text{int}} = \gamma\rho^{\alpha}_{m}e^{-\beta\phi}, \quad p_{\text{int}} = (\alpha-1)\rho_{\text{int}}.
    \label{Eint}
\end{eqnarray}
From Eq. (\ref{mEq}), it can be demonstrated that \(\rho_{m} \propto \frac{1}{a^{3}}\), leading to \(\rho_{m} = \rho_{m_{0}}\left(\frac{a_0}{a^3}\right)\), where $a_0$ and $\rho_{m_{0}}$ is the initial value of the scale factor and the matter energy density, respectively. Throughout this paper, we consider $a_0 = 1$. For the case of closed FLRW spacetime, one can do a similar parametrization as done above and get a similar expression of $\rho_{\text{int}}$ and $p_{\text{int}}$ as shown in Eq.~(\ref{Eint}). We also assume the potential of the scalar field as \(V(\phi) = V_{0}e^{-\lambda\phi}\) for Quintessence-like and Phantom-like scalar fields.

Now, for quintessence like scalar field, we take $\epsilon=1$,
Now for $k=1$,
\begin{figure*}\label{GammavariationQ}
\subfigure[Variation of $a/a_{max}$ with $t/t_{max}$]
{\includegraphics[width=83mm,height=58mm]{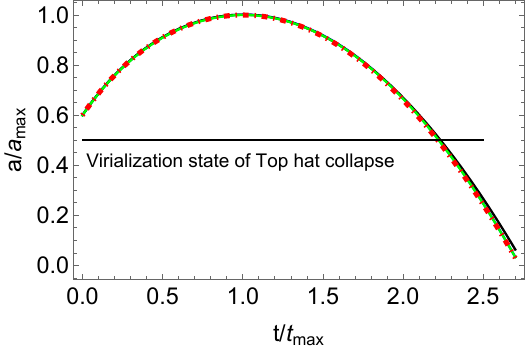}\label{Gammaa}}
\hspace{0.2cm}
\subfigure[Variation of $\omega_{\phi}$ with $t/t_{max}$]
{\includegraphics[width=83mm,height=58mm]{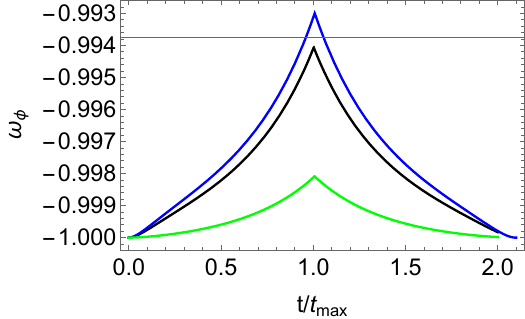}\label{Gammab}}
\hspace{0.2cm}
\subfigure[Variation of $\frac{\rho_{\phi}}{\Bar{\rho}_{\phi}}$ with $t/t_{max}$]
{\includegraphics[width=83mm,height=58mm]{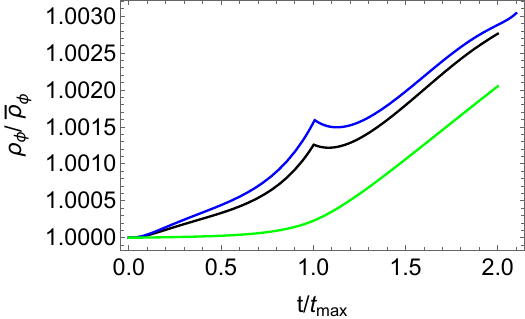}\label{Gammac}}
\hspace{0.2cm}
\subfigure[Variation of $\omega_t$ with $t/t_{max}$ ]
{\includegraphics[width=83mm,height=58mm]{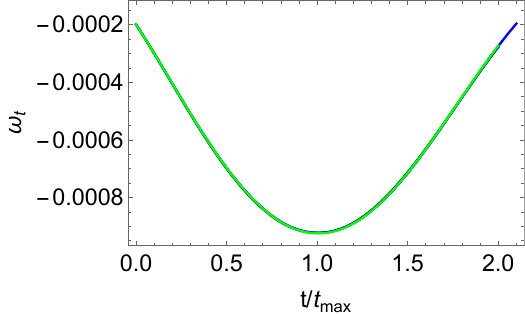}\label{Gammad}}
\hspace{0.2cm}
\subfigure[Variation of $\bar{\omega}_t$ with $t/t_{max}$ ]
{\includegraphics[width=83mm,height=58mm]{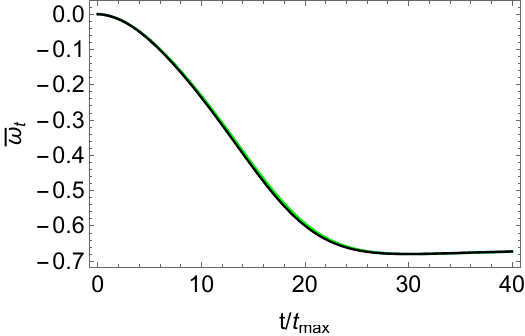}\label{Gammae}}
\hspace{0.2cm}
\caption{Figure shows variation of different variables with variation of $\gamma$ for scalar potential $V(\phi)=V_{0}e^{-\lambda\phi}$ for Quintessence field. Where $a_{max}$ is the maximum value of scale factor and $t_{max}$ is the time when it will reach that value. $V_0$ has the dimension of inverse length squared. Discussion on the unit of $V_0$ can be found in the main text. Here Top hat collapse is represented by red dotted line whereas for the green curves interaction is zero, so it represent minimal coupling. for blue curves $\gamma=.0006$ and for black curves $=.0005$. Here $\beta=1$ is fixed.}
\label{Gamma}
\end{figure*}
\begin{figure*}\label{BetavariationQ}
\subfigure[Variation of $a/a_{max}$ with $t/t_{max}$]
{\includegraphics[width=82mm,height=54mm]{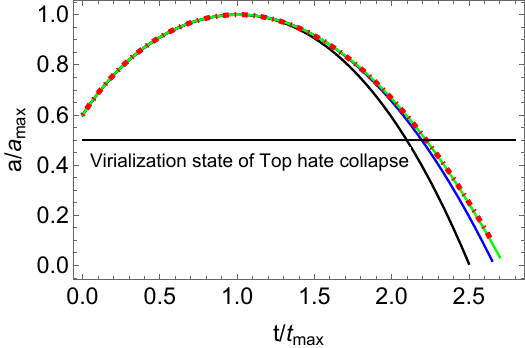}\label{Betaa}}
\hspace{0.2cm}
\subfigure[Variation of $\omega_{\phi}$ with $t/t_{max}$]
{\includegraphics[width=82mm,height=54mm]{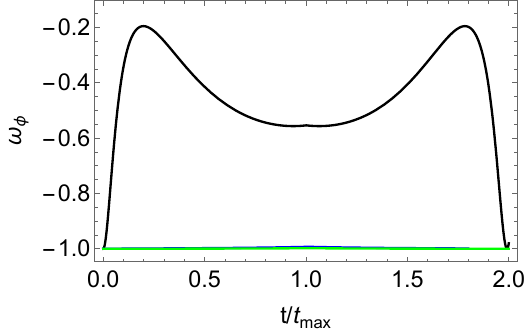}\label{Betab}}
\hspace{0.2cm}
\subfigure[Variation of $\frac{\rho_{\phi}}{\Bar{\rho}_{\phi}}$ with $t/t_{max}$]
{\includegraphics[width=82mm,height=54mm]{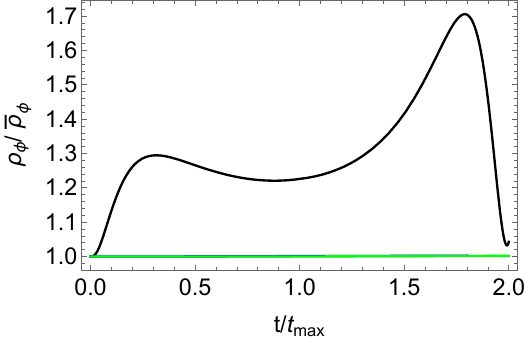}\label{Betac}}
\hspace{0.2cm}
\subfigure[Variation of $\omega_t$ with $t/t_{max}$ ]
{\includegraphics[width=85mm,height=54mm]{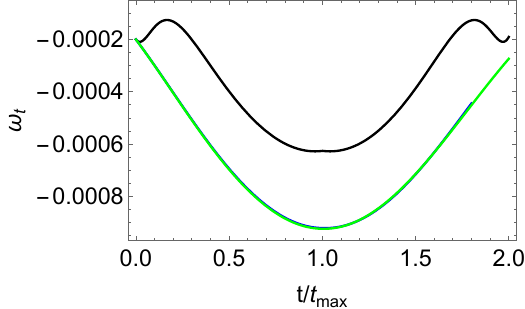}\label{Betad}}
\hspace{0.2cm}
\subfigure[Variation of $\bar{\omega}_t$ with $t/t_{max}$ ]
{\includegraphics[width=85mm,height=54mm]{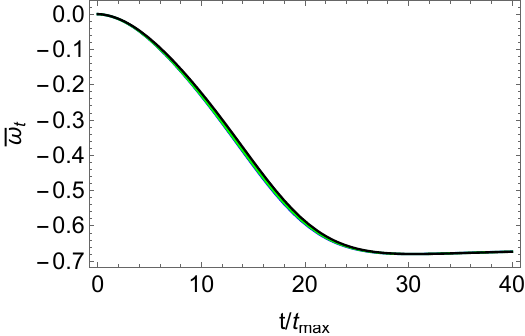}\label{Betae}}
\hspace{0.2cm}
\caption{Figure shows variation of different variables with variation of $\beta$ for scalar potential $V(\phi)=V_{0}e^{-\lambda\phi}$ for Quintessence field. Where $a_{max}$ is the maximum value of the scale factor and $t_{max}$ is the time when it will reach that value. $V_0$ has the dimension of inverse length squared. Discussion on the unit of $V_0$ can be found in the main text. Here Top hat collapse is represented by a red dotted line whereas for the green curves interaction is zero, so it represents minimal coupling. For blue curves $\beta=1$ and for black curves $\beta=100$, and for both cases $\gamma=.0006$ is fixed.}
\label{Beta}
\end{figure*}
\begin{figure*}\label{AlphavariationQ}
\subfigure[Variation of $a/a_{max}$ with $t/t_{max}$]
{\includegraphics[width=82mm,height=58mm]{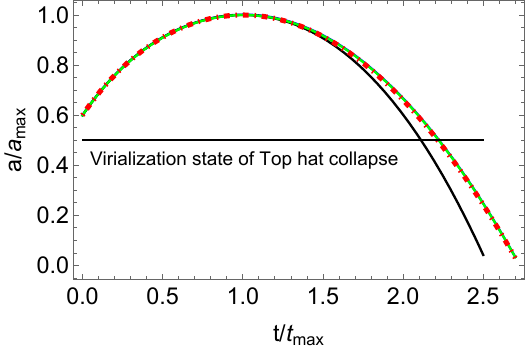}\label{Alphaa}}
\hspace{0.2cm}
\subfigure[Variation of $\omega_{\phi}$ with $t/t_{max}$]
{\includegraphics[width=82mm,height=58mm]{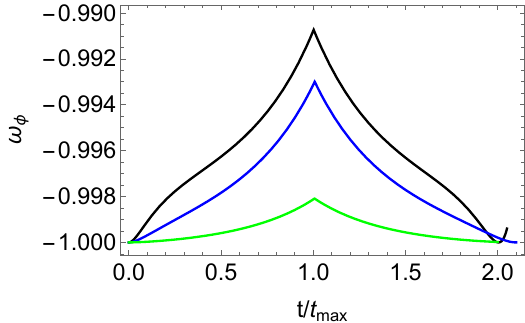}\label{Alphab}}
\hspace{0.2cm}
\subfigure[Variation of $\frac{\rho_{\phi}}{\Bar{\rho}_{\phi}}$ with $t/t_{max}$]
{\includegraphics[width=82mm,height=58mm]{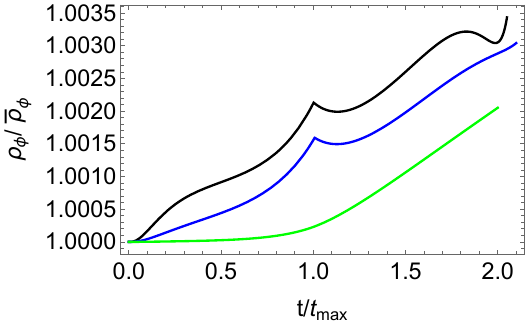}\label{Alphac}}
\hspace{0.2cm}
\subfigure[Variation of $\omega_t$ with $t/t_{max}$ ]
{\includegraphics[width=82mm,height=58mm]{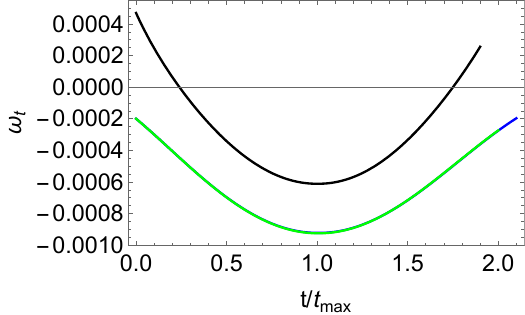}\label{Alphad}}
\hspace{0.2cm}
\subfigure[Variation of $\bar{\omega}_t$ with $t/t_{max}$ ]
{\includegraphics[width=82mm,height=58mm]{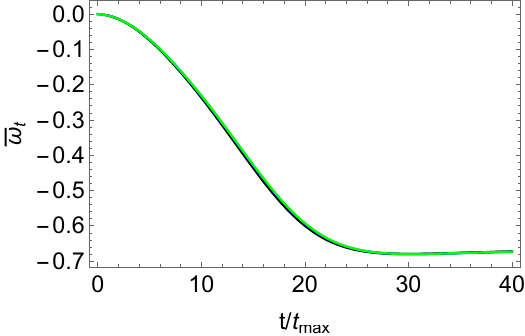}\label{Alphae}}
\hspace{0.2cm}
\caption{Figure shows variation of different variables with variation of $\alpha$ for scalar potential $V(\phi)=V_{0}e^{-\lambda\phi}$ for Quintessence field. Where $a_{max}$ is the maximum value of scale factor and $t_{max}$ is the time when it will reach that value. $V_0$ has the dimension of inverse length squared. Discussion on the unit of $V_0$ can be found in the main text. Here Top hat collapse is represented by red dotted line whereas for the green curves interaction is zero, so it represent minimal coupling. For blue curves $\alpha=1$ and for black curves $\alpha=1.5$, and for both cases $\gamma=.0006$ is fixed.}
\label{Alpha}
\end{figure*}
\begin{figure*}\label{BetavariationPh}
\subfigure[Variation of $a/a_{max}$ with $t/t_{max}$]
{\includegraphics[width=82mm,height=54mm]{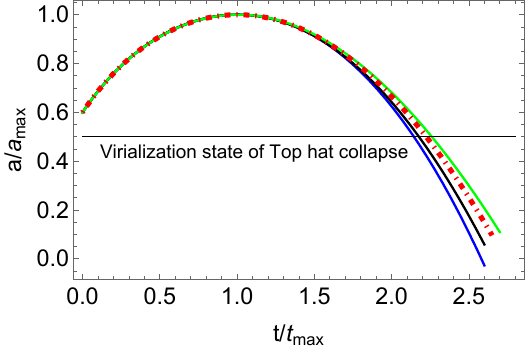}\label{PhBetaa}}
\hspace{0.2cm}
\subfigure[Variation of $\omega_{\phi}$ with $t/t_{max}$]
{\includegraphics[width=82mm,height=54mm]{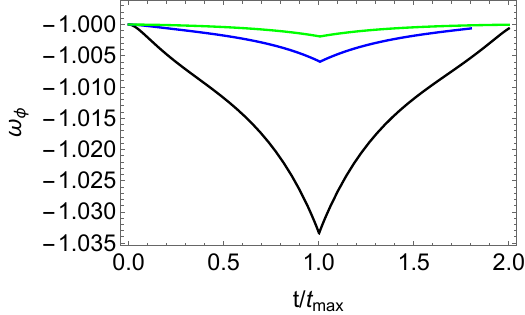}\label{PhBetab}}
\hspace{0.2cm}
\subfigure[Variation of $\frac{\rho_{\phi}}{\Bar{\rho}_{\phi}}$ with $t/t_{max}$]
{\includegraphics[width=82mm,height=54mm]{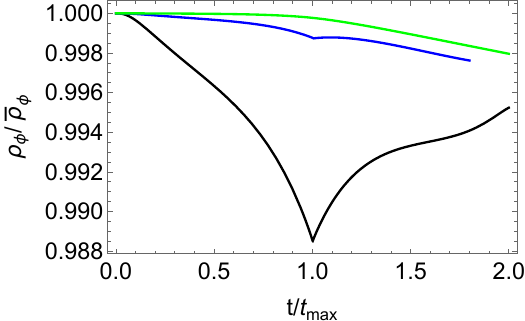}\label{PhBetac}}
\hspace{0.2cm}
\subfigure[Variation of $\omega_t$ with $t/t_{max}$ ]
{\includegraphics[width=85mm,height=54mm]{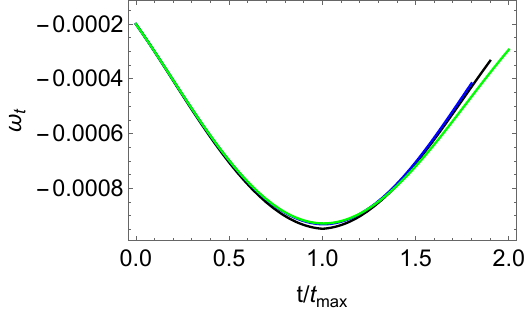}\label{PhBetad}}
\hspace{0.2cm}
\subfigure[Variation of $\bar{\omega}_t$ with $t/t_{max}$ ]
{\includegraphics[width=85mm,height=54mm]{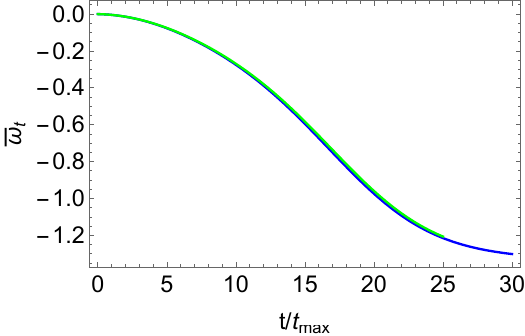}\label{PhBetae}}
\hspace{0.2cm}
\caption{Figure shows variation of different variables with variation of $\beta$ for scalar potential $V(\phi)=V_{0}e^{-\lambda\phi}$ for Phantom field. Where $a_{max}$ is the maximum value of scale factor and $t_{max}$ is the time when it will reach that value. $V_0$ has the dimension of inverse length squared. Discussion on the unit of $V_0$ can be found in the main text. Here Top hat collapse is represented by red dotted line whereas for the green curves interaction is zero, so it represent minimal coupling. For blue curves $\beta=5$ and for black curves $\beta=20$, and for both cases $\gamma=.0001$ is fixed.}
\label{PhBeta}
\end{figure*}
\begin{figure*}\label{GammavariationPh}
\subfigure[Variation of $a/a_{max}$ with $t/t_{max}$]
{\includegraphics[width=82mm,height=58mm]{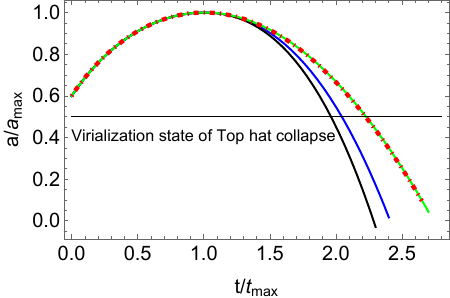}\label{PhGammaa}}
\hspace{0.2cm}
\subfigure[Variation of $\omega_{\phi}$ with $t/t_{max}$]
{\includegraphics[width=82mm,height=58mm]{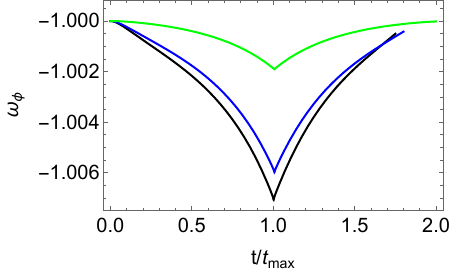}\label{PhGammab}}
\hspace{0.2cm}
\subfigure[Variation of $\frac{\rho_{\phi}}{\Bar{\rho}_{\phi}}$ with $t/t_{max}$]
{\includegraphics[width=82mm,height=58mm]{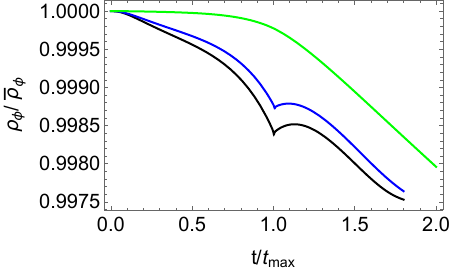}\label{PhGammac}}
\hspace{0.2cm}
\subfigure[Variation of $\omega_t$ with $t/t_{max}$ ]
{\includegraphics[width=85mm,height=58mm]{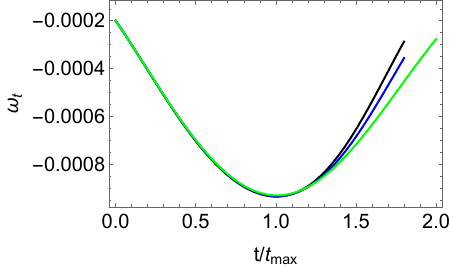}\label{PhGammad}}
\hspace{0.2cm}
\subfigure[Variation of $\bar{\omega}_t$ with $t/t_{max}$ ]
{\includegraphics[width=85mm,height=58mm]{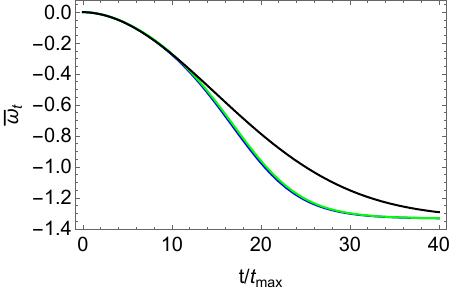}\label{PhGammae}}
\hspace{0.2cm}
\caption{Figure shows variation of different variables with variation of $\gamma$ for scalar potential $V(\phi)=V_{0}e^{-\lambda\phi}$ for Phantom field. Where $a_{max}$ is the maximum value of scale factor and $t_{max}$ is the time when it will reach that value. $V_0$ has the dimension of inverse length squared. Discussion on the unit of $V_0$ can be found in the main text. Here Top hat collapse is represented by red dotted line whereas for the green curves interaction is zero, so it represent minimal coupling. For blue curves $\gamma=.0005$ and for black curves $\gamma=.0006$, and for both cases $\beta=1$ is fixed.}
\label{PhGamma}
\end{figure*}
\begin{figure*}\label{AlphavariationPh}
\subfigure[Variation of $a/a_{max}$ with $t/t_{max}$]
{\includegraphics[width=82mm,height=54mm]{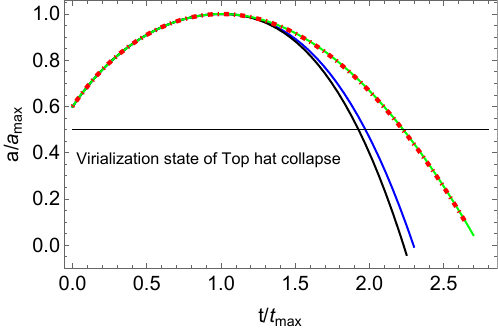}\label{PhAlphaa}}
\hspace{0.2cm}
\subfigure[Variation of $\omega_{\phi}$ with $t/t_{max}$]
{\includegraphics[width=82mm,height=54mm]{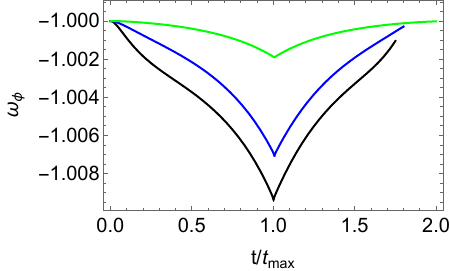}\label{PhAlphab}}
\hspace{0.2cm}
\subfigure[Variation of $\frac{\rho_{\phi}}{\Bar{\rho}_{\phi}}$ with $t/t_{max}$]
{\includegraphics[width=82mm,height=54mm]{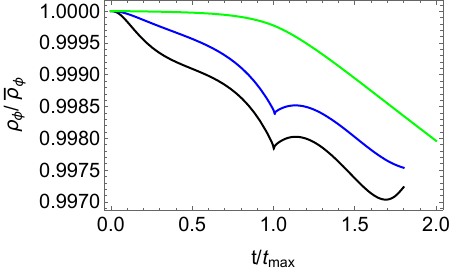}\label{PhAlphac}}
\hspace{0.2cm}
\subfigure[Variation of $\omega_t$ with $t/t_{max}$ ]
{\includegraphics[width=85mm,height=54mm]{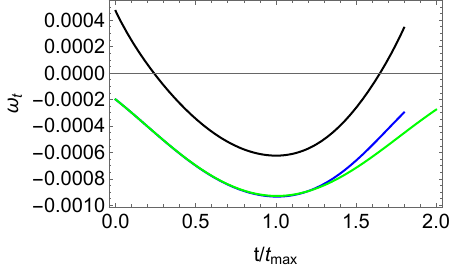}\label{PhAlphad}}
\hspace{0.2cm}
\subfigure[Variation of $\bar{\omega}_t$ with $t/t_{max}$ ]
{\includegraphics[width=85mm,height=54mm]{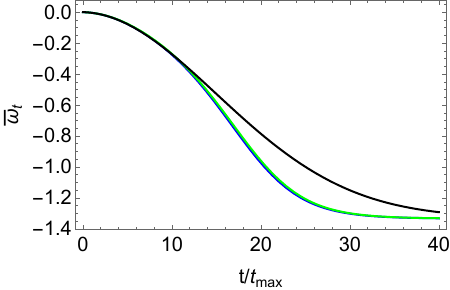}\label{PhAlphae}}
\hspace{0.2cm}
\caption{Figure shows variation of different variables with variation of $\alpha$ for scalar potential $V(\phi)=V_{0}e^{-\lambda\phi}$ for Phantom field. Where $a_{max}$ is the maximum value of scale factor and $t_{max}$ is the time when it will reach that value. $V_0$ has the dimension of inverse length squared. Discussion on the unit of $V_0$ can be found in the main text. Here Top hat collapse is represented by red dotted line whereas for the green curves interaction is zero, so it represent minimal coupling. For blue curves $\alpha=1$ and for black curves $\alpha=1.5$, and for both cases $\beta=1$ and $\gamma=.0006$ is fixed.}
\label{PhAlpha}
\end{figure*}
Substituting these expressions in Eq.~(\ref{eleven}), one can get the following second-order differential equation of $\phi(a)$:
\begin{widetext}
\begin{eqnarray}\label{equationforQk=1}
&-&4V_{0}e^{-\lambda\phi}\phi_{,a}a^{3}+9\phi_{,a}a+\frac{V_{0}e^{-\lambda\phi}\phi_{,a}^{3}a^{5}}{2}+\frac{\rho_{m_{0}}\phi_{,a}^{3}a^{2}}{4}-\phi_{,a}^{3}a^{3}+3\lambda a^{2}V_{0}e^{-\lambda\phi}-\frac{5\rho_{m_{0}}\phi_{,a}}{2}-\rho_{m_{0}}a\phi_{,aa}+3a^{2}\phi_{,aa}-V_{0}e^{-\lambda\phi}a^{4}\phi_{,aa}\nonumber\\&-&\frac{\lambda V_{0}e^{-\lambda\phi}\phi_{,a}^{2}a^{4}}{2}+\frac{\gamma(\rho_{m_{0}})^{\alpha}e^{-\beta\phi}}{a^{3\alpha}}\left[3\beta a^{2}-\frac{\beta\phi_{,a}^{2}a^{4}}{2}-4\phi_{,a}a^{3}+\frac{\phi^{3}_{,a}a^{5}}{2}-\frac{\alpha\phi^{3}_{,a}a^{5}}{4}+\frac{3\alpha\phi_{,a}a^{3}}{2}-\phi_{,aa}a^{4}\right]=0\, ,
\end{eqnarray}
\end{widetext}
and for $k=0$, we get
\begin{widetext}
\begin{eqnarray}
\label{equationforQk=0}
&-&4V_{0}e^{-\lambda\phi}\phi_{,a}a^{3}+\frac{V_{0}e^{-\lambda\phi}\phi_{,a}^{3}a^{5}}{2}+\frac{\rho_{m_{0}}\phi_{,a}^{3}a^{2}}{4}+3\lambda a^{2}V_{0}e^{-\lambda\phi}-\frac{5\rho_{m_{0}}\phi_{,a}}{2}-\rho_{m_{0}}a\phi_{,aa}-V_{0}e^{-\lambda\phi}a^{4}\phi_{,aa}\nonumber\\&-&\frac{\lambda V_{0}e^{-\lambda\phi}\phi_{,a}^{2}a^{4}}{2}+\frac{\gamma(\rho_{m_{0}})^{\alpha}e^{-\beta\phi}}{a^{3\alpha}}\left[3\beta a^{2}-\frac{\beta\phi_{,a}^{2}a^{4}}{2}-4\phi_{,a}a^{3}+\frac{\phi^{3}_{,a}a^{5}}{2}-\frac{\alpha\phi^{3}_{,a}a^{5}}{4}+\frac{3\alpha\phi_{,a}a^{3}}{2}-\phi_{,aa}a^{4}\right]=0
\end{eqnarray}
\end{widetext}
It is generally considered that the phantom-like scalar
field has negative kinetic energy and therefore, for the
phantom field $\epsilon=-1$. For the phantom field, Substituting these expressions in Eq.~(\ref{eleven}), one can get the following second-order differential equation of $\phi(a)$:
\begin{widetext}
\begin{eqnarray}\label{equationforQk=1}
&&4V_{0}e^{-\lambda\phi}\phi_{,a}a^{3}-9\phi_{,a}a+\frac{V_{0}e^{-\lambda\phi}\phi_{,a}^{3}a^{5}}{2}+\frac{\rho_{m_{0}}\phi_{,a}^{3}a^{2}}{4}-\phi_{,a}^{3}a^{3}+3\lambda a^{2}V_{0}e^{-\lambda\phi}+\frac{5\rho_{m_{0}}\phi_{,a}}{2}+\rho_{m_{0}}a\phi_{,aa}-3a^{2}\phi_{,aa}+V_{0}e^{-\lambda\phi}a^{4}\phi_{,aa}\nonumber\\&+&\frac{\lambda V_{0}e^{-\lambda\phi}\phi_{,a}^{2}a^{4}}{2}+\frac{\gamma(\rho_{m_{0}})^{\alpha}e^{-\beta\phi}}{a^{3\alpha}}\left[3\beta a^{2}+\frac{\beta\phi_{,a}^{2}a^{4}}{2}+4\phi_{,a}a^{3}-\frac{\phi^{3}_{,a}a^{5}}{2}-\frac{\alpha\phi^{3}_{,a}a^{5}}{4}-\frac{3\alpha\phi_{,a}a^{3}}{2}+\phi_{,aa}a^{4}\right]=0\, ,
\end{eqnarray}
\end{widetext}
and for $k=0$, we get
\begin{widetext}
\begin{eqnarray}\label{equationforQk=0}
&&4V_{0}e^{-\lambda\phi}\phi_{,a}a^{3}+\frac{V_{0}e^{-\lambda\phi}\phi_{,a}^{3}a^{5}}{2}+\frac{\rho_{m_{0}}\phi_{,a}^{3}a^{2}}{4}+3\lambda a^{2}V_{0}e^{-\lambda\phi}+\frac{5\rho_{m_{0}}\phi_{,a}}{2}+\rho_{m_{0}}a\phi_{,aa}+V_{0}e^{-\lambda\phi}a^{4}\phi_{,aa}\nonumber\\&+&\frac{\lambda V_{0}e^{-\lambda\phi}\phi_{,a}^{2}a^{4}}{2}+\frac{\gamma(\rho_{m_{0}})^{\alpha}e^{-\beta\phi}}{a^{3\alpha}}\left[3\beta a^{2}+\frac{\beta\phi_{,a}^{2}a^{4}}{2}+4\phi_{,a}a^{3}-\frac{\phi^{3}_{,a}a^{5}}{2}-\frac{\alpha\phi^{3}_{,a}a^{5}}{4}-\frac{3\alpha\phi_{,a}a^{3}}{2}+\phi_{,aa}a^{4}\right]=0\, ,
\end{eqnarray}
\end{widetext}

Where $\phi_{,aa}$ is the second-order derivative of the scalar
field with respect to $a$. Here the last terms involving factor $\gamma$ is coming due to the non-minimal interaction between matter and scalar field. If we put $\gamma=0$ in this equation then we will get the equation where there is no coupling between scalar field and dark matter.
We can now solve the above differential equation to get the functional form of $\phi(a)$. Consequently, using the solution of $\phi(a)$ and the differential Eqs. (\ref{three}), (\ref{four}), we can obtain the expression of scale factor $a$ as a function of co-moving time $t$. In the next section, we discuss the results of our work elaborately.
\section{Results obtained from the collapsing process}

In our prior study of structure formation of dark matter in the presence of a minimally coupled scalar field \cite{Saha1}, we identified a bounded region in a region plot that links the initial value of the dark matter, $\rho_{m_{0}}$, with the scalar field potential parameter $V_{0}$. The specific values of $\rho_{m_{0}}$ and $V_{0}$ within this region are indicative of the dynamics exhibited by overdense regions, characterized by a collapsing phase following an initial expansion akin to the top-hat collapse. The allowed parameter space for $\rho_{m_{0}}$ and $V_{0}$ reveals that for larger values of $V_{0}$, $\rho_{m_{0}}$ must take smaller values, and vice versa, in order to achieve dynamics similar to the top-hat collapse. In this paper, we conduct a comparable study for non-minimal scenarios and our results reveal a distinct contrast from the preceding minimal scenario. Figures (\ref{Fig1}, \ref{Fig2}, \ref{Fig3}, \ref{Fig4}) depict the allowed shaded region of $\rho_{m_{0}}$ and $V_{0}$ associated with dynamics resembling top-hat collapse when non-minimal couplings are taken into account, where we consider different values of $\alpha$ for non-minimally coupled quintessence-like (i.e., Figs.~(\ref{Fig1}, \ref{Fig2})) and phantom-like scalar fields (i.e., Figs.~(\ref{Fig3}, \ref{Fig4})). On the other hand, the unshaded regions in those figures correspond to those dynamics that expand eternally. The region plots demonstrate that while the allowed region for $\rho_{m_{0}}$ and $V_{0}$ remains similar to that of the minimal coupling scenario for lower values of $\alpha$, its characteristics undergo a change for higher values of the same parameter. From Figs.~(\ref{Fig2}, \ref{Fig4}), it can be seen that for $\alpha = 7$, there exists a range of values of $V_0$ for which two allowed ranges of $\rho_{m_{0}}$ are possible. However, it is also noteworthy that there exists a range of $V_0$ for which all values of $\rho_{m_{0}}$ are permitted. This type of nature is absent if we consider a minimally coupled scalar field. Based on the aforementioned findings, it can be broadly asserted that the inclusion of non-minimal coupling expands the parameter space of $\rho_{m_{0}}$ and $V_{0}$ in which overdense regions exhibit initial expansion followed by subsequent contraction. Whereas the region for continual expansion decreases due to the non-minimal coupling. It's important to highlight that the features of the allowed regions are not as strongly influenced by variations in the values of other non-minimal coupling parameters, specifically $\gamma$ and $\beta$, as they are by $\alpha$. However, for all cases, as stated before, the shaded region increases as we increase the value of any of the non-minimal coupling parameters. 

Now, considering some of the values of $\rho_{m_{0}}$ and $V_{0}$ corresponding to the shaded regions, we solve the second order differential Eqs.~(\ref{equationforQk=1}),(\ref{equationforQk=0}) of $\phi(a)$. Since the differential equations, Eqs.~(\ref{equationforQk=1}) and (\ref{equationforQk=0}), are second-order differential equations, we must consider two initial conditions, namely $\phi(a=1)$ and $\phi^{\prime}(a=1),$ for their solution. These initial values are selected to ensure that both $\dot{a}$ and $p_\phi/\rho_\phi,$ which depend on both $\phi$ and $\phi^{\prime},$ attain reasonable values at the initial point when the collapse begins. Specifically, the chosen initial values are such that $p_\phi/\rho_\phi$ is approximately $-1$, allowing the scalar field sector to function as a dark energy component. To facilitate a comparison between our model and the standard top-hat collapse model, we restrict our discussion in this paper to scenarios where the initial value of $\dot a$ is positive. This positive value ensures an initial expansion phase of the over-dense region, aligning with the conditions of the standard top-hat collapse model. In our calculations, we consider the initial conditions $\phi(a=1) = 0.001$ and $\phi^{\prime}(a=1) = 0.00001$ for solving the differential equations (Eqs.~(\ref{equationforQk=1}), (\ref{equationforQk=0})). Here, we deal with six parameters in our model: $V_{0}$, $\rho_{m_{0}}$, $\lambda$, $\alpha$, $\beta$, and $\gamma$. In our system of geometrical units, the scalar field, $\lambda$, $\alpha$, and $\beta$ are dimensionless, while $a$ has the dimension of length, $\rho_{m_0}$ and $V_0$ have dimensions of inverse length squared and $\gamma$ has dimension $L^{2(\alpha -1)}$, where $L$ represents the length dimension.
Similar to our approach in the previous study with minimal coupling, in the current investigation, we choose values of $\rho_{m_0}$ that consistently remain proximate to the critical density of the background at the onset of the collapse. For a specific epoch where collapse occurs, if we express $\rho_{m_0}$ in conventional units, we can easily convert it into geometrized units by multiplying $\rho_{m_0}$ by $Gc^{-4}$, where $G$ is the universal gravitational constant and $c$ is the velocity of light. This conversion yields a value of $L^{-2}$. This value in units of $L^{-2}$ serves as a suitable scale in our context to express the values of $V_{0}$, $\rho_{m_{0}}$, and $\gamma$.

Now, to investigate the impact of non-minimal coupling on the evolution of overdense regions, we set the values of $\lambda$, $\rho_{m_{0}}$, and $V_{0}$ to $\lambda=1$, $\rho_{m_{0}}=5$, and $V_{0}=0.001$, respectively. We then vary one of the coupling parameters, $\alpha$, $\beta$, or $\gamma$, while keeping the values of the remaining two fixed. Figs.~ (\ref{Gamma}), (\ref{Beta}), and (\ref{Alpha}) depict the evolution of dynamic quantities, such as $a$, $\omega_{\phi}=p_\phi/\rho_\phi$, $\delta_\phi = \rho_\phi/\bar{\rho}_\phi$, $\omega_t=p_\phi/(\rho_m + \rho_\phi)$, and $\overline{\omega}_t=\overline{p}_{\phi}/(\overline{\rho}_m + \overline{\rho}_\phi)$, with time for different values of $\gamma$, $\beta$, and $\alpha$, respectively. Here, $\omega_{\phi}$ represents the equation of state of the quintessence-like scalar field in the overdense region, $\omega_t$ denotes the effective or total equation of state of the internal fluid comprising matter and the quintessence-like scalar field, and $\overline{\omega}_t$ represents the total equation of state of the background fluid consisting of matter and the quintessence-like scalar field. On the other hand, Figs.~ (\ref{PhGamma}), (\ref{PhBeta}), and (\ref{PhAlpha}) illustrate the evolution of $a$, $\omega_{\phi}$, $\delta_\phi$, $\omega_t$, and $\overline{\omega}_t$ when the scalar field exhibits phantom-like nature. The evolution of the scale factor $a$ (as depicted in Figs.~(\ref{Gammaa}), (\ref{Betaa}), (\ref{Alphaa}), (\ref{PhGammaa}), (\ref{PhBetaa}), (\ref{PhAlphaa})) reveals that non-minimal coupling accelerates the over-dense region's transition to the virialization state compared to both minimal coupling scenarios and the top-hat collapse. Whereas the nature of the evolution of $\omega_t$ illustrated in Figs.~(\ref{Gammad}), (\ref{Betad}), (\ref{Alphad}), (\ref{PhGammad}), (\ref{PhBetad}), (\ref{PhAlphad})) suggests that as we increase the values of non-minimal coupling parameters, the resulting fluid begins to exhibit behavior more akin to that of dust. 
The reason behind this lies in the fact that, as demonstrated, the energy density of matter remains unaffected by the non-minimal interaction, staying proportional to $\frac{1}{a(t)^3}$. However, the Klein-Gordon equation of the scalar field undergoes modification with an additional interaction term (Eq.(\ref{Klein1})). Consequently, the presence of dark matter slows down the flux of dark energy through the boundary, causing less pressure at the boundary and throughout the over-dense region. As we are aware, the expression for pressure is given by $p= -\frac{\dot{F}(r,t)}{\dot{R}R^2}$. This implies that when the pressure is zero, it corresponds to a Misner-Sharp mass $F$ that is independent of time. Therefore, in the scenario of negligible pressure, we can express $F$ as $F = F_0 r^3 + \delta F_0(t) r^3$, where $\delta F_0(t)\to 0$ and $F_0$ is a positive valued constant. Since at the boundary $F(r_b, t)= 2M(r_v,v)$ (Eq.~(\ref{FM1})), the total internal mass measured from the external Vaidya spacetime becomes nearly time-independent. This suggests that the internal system behaves almost like an isolated universe, akin to the top-hat collapse model. Additionally, contemplating larger values for the non-minimal coupling parameters facilitates the transition of dark energy from its homogeneous state, causing it to cluster more within the over-dense region. This phenomenon is evident in the evolution of $\delta_\phi = \rho_\phi/\bar{\rho}_\phi$ as depicted in Figs.~(\ref{Gammac}), (\ref{Betac}), (\ref{Alphac}), (\ref{PhGammac}), (\ref{PhBetac}), (\ref{PhAlphac}).

\section{Discussion and conclusion}
\label{sec4}
In this paper, we investigate how a non-minimally coupled scalar field influences the evolution of the over-dense region of dark matter. We adopt a spacetime structure crucial for preserving the homogeneous behavior of dark energy, as seen in a previous study \cite{Saha1}. Our focus is on algebraic coupling, where the interaction Lagrangian is independent of derivatives of the scalar field. We explore the impact of non-minimal coupling on the virialized structures of dark matter, particularly on a cosmological scale where dark energy's influence cannot be disregarded.
The motivation for exploring non-minimal coupling comes from notable deviations observed in earlier results with minimal coupling when compared to the standard top-hat collapse model. Moreover non-minimal coupling in the dark sector cannot be ruled out in principle \cite{Hussain:2023kwk, Hussain:2022dhp, Bhattacharya:2022wzu, Hussain:2022osn, Chatterjee:2021ijw}. As because we do not know exactly the components of the dark sector and there exists no thermodynamic rule to prevent energy-momentum exchange in the dark sector, in general one can always assume some non-minimal coupling. These couplings will have interesting cosmological consequences which have been reported in the previous references. In the present work we have implemented the idea of non-minimal coupling in the dark sector in the level of the action and worked out the whole theory. The effective nature of the coupling although remains phenomenological, as shown in Eq.~(\ref{Eint}). One can choose various forms of this interaction term, out of various possibilities we have chosen one that keeps the dynamics tractable and simple. The question of the exact form of this interaction term remains open as it cannot be dictated by any formal theory. Our study reveals that increased non-minimal coupling induces the clustering of dark energy within the over-dense region of dark matter and the energy density of matter remains unaffected, being proportional to $\frac{1}{a(t)^3}$. On the other hand, the Klein-Gordon equation of the scalar field undergoes modification with an additional interaction term, influencing dark energy to cluster with dark matter.

In a previous work \cite{Creminelli:2009mu} it was argued that if the dark energy component of the universe is modelled by a scalar field whose Lagrangian density contains a nontrivial function of the kinetic term, then in those cases the sound speed in the dark energy sector can be really small. In such cases, the dark energy sector can non-trivially affect structure formation. In our case, the dark energy component is modelled by a scalar field which has the standard kinetic term but still affects the structure formation process non-trivially. This happens due to the non-minimal coupling in the dark sector. Due to the specific non-minimal coupling used in our work we see that although in the expanding, flat background FLRW spacetime the equation of state is effectively like dark energy the effective equation of state of matter in the detached collapsing spacetime is approximately equal to the equation of state of dust. Gravitational contraction in a closed FLRW spacetime with non-minimal coupling in the matter components produce such a state. As a result of this the gravitational collapse in non-minimally coupled dark sector becomes a bit more easy to handle than the case where the components are not coupled. In the simplest case the external matched spacetime just turns out to be the Schwarzschild spacetime when the equation of state of the collapsing matter practically nears zero. As because in this case the internal spacetime has a dust like effective component, the whole collapse process conserves mass.

In this paper we have used general relativistic paradigm to formulate the collapsing process. This was a necessity as we have started at the level of the action which has the gravitational as well as the matter and non-minimal coupling terms. The minimization of the action produced all the equations of motion which we have used in this paper. Inspite of this formal approach our work has a heavy phenomenological flavour as because the form of the non-minimal coupling and the scalar field potential were chosen from the widely used forms by various authors working in this field. These forms can reasonably reproduce the late-time cosmology results. Our semi-formal approach ends near virialization. Our work does not analytically predict virialization but virialization is also phenomenologically implanted. This feature is not new, the concept of virialization at the end phase of top hat collapse was also implemented in the same manner. Our work predicts that this final virialized form of the effective matter inside the detached spherical overdensity is rather exotic. This effective matter is a combination of the dark energy component and the pressure-less dark matter component. Both the sectors are exchanging energy and momentum in such a way that the effective matter also behaves almost as dust. This prediction can have interesting consequences in theories of structure formation. Our work predicts that the essential input for clusters of galaxies may have dark energy components hidden somewhere. We will like to explore this topic in the near future.

We extensively explore the permissible parameter space of $V_0$ and $\rho_{m_0}$, identifying conditions under which overdense regions demonstrate dynamics akin to top-hat collapse which has an initial expansion phase followed by a collapsing phase culminating in the virialization state. The study encompasses both quintessence-like and phantom-like scenarios, revealing distinct changes in the parameter space for non-minimal coupling compared to minimal coupling. We show that the allowed parameter space increases as we increase the values of non-minimal coupling parameters.
Additionally, we study the impact of non-minimal coupling on the virialization process. We present the evolution of dynamic quantities, such as $a$, $\omega_{\phi}$, $\delta_\phi$, $\omega_t$, and $\overline{\omega}_t$, for different values of the non-minimal coupling parameters. Our results imply that the non-minimal coupling accelerates the transition to the virialization state and leads to behavior resembling that of dust in the resulting fluid.
Furthermore, we highlight the nearly time-independent behavior of the total internal mass in the presence of negligible pressure, implying the internal system behaves similarly to an isolated universe, as in the top-hat collapse model. Larger values of non-minimal coupling parameters facilitate the transition of dark energy from its homogeneous state, causing increased clustering within the over-dense region.
Our findings have the potential to enhance our comprehension of the cosmological implications associated with non-minimal coupling of dark matter and dark energy in the context of dark matter structure formation, along with its possible observational signatures.

\section{Acknowledgement}
DD would like to acknowledge the support of
the Atlantic Association for Research in the Mathematical Sciences (AARMS) for funding the work.

\end{document}